\documentclass[hyper,11pt,paper]{article}
\usepackage[centertags]{amsmath}
\usepackage{amssymb}
\usepackage{graphicx}
\usepackage{epsfig}
\usepackage{ulem}
\usepackage[english]{babel}
\usepackage{array}
\usepackage{amsthm}
\usepackage{latexsym}
\usepackage[mathcal]{euscript}
\pdfoutput=1
\usepackage{epsfig}
 \usepackage{jheppub}
\usepackage{mathdots}
\usepackage{MnSymbol}
\usepackage{physics}
\usepackage{multirow}
\usepackage{ dsfont }
\usepackage{capt-of}
\usepackage{caption}
\usepackage{subcaption}
\usepackage{soul}
\usepackage{comment}

\usepackage{tikz}
\usetikzlibrary{calc,decorations.pathmorphing,arrows.meta,fadings,decorations.markings}
\tikzset{>={Stealth[width=2mm,length=2mm]}}

\newcommand{\cB}{\mathcal{B}}
\newcommand{\cC}{\mathcal{C}}

\newcommand{\cM}{\mathcal{M}}

\newcommand{\cO}{\mathcal{O}}

\newcommand{\cS}{\mathcal{S}}

\definecolor{dgreen}{rgb}{0.,0.6,0.}

\usepackage{scalerel,stackengine}

\newcommand{\be}{\begin{equation}}
\newcommand{\ee}{\end{equation}}
\newcommand{\ba}{\begin{eqnarray}}
\newcommand{\ea}{\end{eqnarray}}
\newcommand{\bea}{\begin{equation}\begin{aligned}}
\newcommand{\eea}{\end{aligned}\end{equation}}

\newcommand{\dslash}{\delta^{\!\!\!\!-}\!}
\newcommand{\mn}{\mu\nu}
\newcommand{\ads}{\text{AdS}}

\title{Quantum corrected black hole microstates and entropy}

\author[a]{Dongming He,}
\author[a]{Juan Hernandez,}
\author[a]{and Maria Knysh}
\affiliation[a]{Theoretische Natuurkunde, Vrije Universiteit Brussel (VUB) and The International Solvay Institutes, Pleinlaan 2, B-1050 Brussels, Belgium}
\emailAdd{dongming.he@vub.be}
\emailAdd{juan.hernandez@vub.be}
\emailAdd{maria.knysh@vub.be}
\abstract{
We extend the semiclassical black hole microstate construction to include quantum corrections to the microscopic entropy using a doubly holographic model of black holes. Specifically, we consider a double-sided black hole on a JT brane with holographic matter, coupled to a pair of holographic CFTs on the asymptotic boundaries. The dimension of the Hilbert space spanned by the microstates of this doubly holographic black hole is given by the exponentiated entropy, which is equal to the sum of the quantum-corrected thermodynamic entropies of the left and right black holes. Importantly, the quantum-corrected thermodynamic entropy is shown to be equal to the generalised entropy of the eternal black hole, and thus can be interpreted as quantifying the entanglement between the two asymptotic boundaries.
}

\begin{document}

\maketitle
\section{Introduction}
\label{sec: Intro}
A fundamental goal of quantum gravity is to provide a microscopic interpretation for the Bekenstein-Hawking entropy of a black hole, which in the semiclassical limit is the area of the horizon such that $S_{\text{BH}}=\frac{\text{Area}}{4 G_N}$  \cite{Bekenstein:1973ur,Hawking:1975vcx}.
Recently, an explicit construction of a family of semiclassical black hole microstates \cite{Balasubramanian:2022gmo, Balasubramanian:2022lnw} was put forward, which consists of shells of matter hidden behind the horizon of spherical black holes in AdS and flat spacetime.
These microstates have small overlaps due to the contribution of Euclidean wormholes to the gravity path integral, and are shown to span a Hilbert space of dimension $e^{S_{\text{BH}}}$, correctly reproducing the Bekenstein-Hawking entropy of the black hole.
Further developments in this direction include \cite{Chandra:2022fwi,Climent:2024trz,Guo:2024zmr,Boruch:2024kvv,Antonini:2024mci,Geng:2024jmm,Balasubramanian:2024yxk,Balasubramanian:2024rek,Abdalla:2025gzn,Magan:2025hce,Balasubramanian:2025zey,Balasubramanian:2025hns,Balasubramanian:2025akx,Antonini:2025ioh}.

In this work, we extend the microstate construction beyond the semiclassical limit by focusing on black holes coupled to holographic matter and using a doubly holographic model \cite{Grimaldi:2022suv}. 
Specifically, we show that the dimension of the Hilbert space spanned by the black hole microstates is given by the exponential of the black hole entropy, which in this case includes quantum corrections of order $O(G_N^0)$ in addition to the original area term, extending the previous results. 
This model consists of an $\ads_2$ black hole with holographic matter coupled to a pair of holographic CFTs on the asymptotic boundaries.
Using brane-world models of double holography \cite{Randall:1999vf,Gubser:1999vj,Karch:2000ct,Karch:2000gx,Almheiri:2019hni} (string theory examples of double holography include \cite{Coccia:2021lpp,Uhlemann:2021nhu,Karch:2022rvr,He:2024djr}), this system can be described in two other equivalent ways: a pair of CFTs on an Euclidean torus with a conformal defect along $\phi=0$, or a BTZ black hole with a 2d JT brane anchored at two asymptotic boundaries. The JT brane forms an interface in the bulk geometry. 
The three equivalent pictures are called the brane, the boundary and the bulk perspectives. 
An appealing feature of doubly holographic models is that quantum corrections in the brane perspective are captured by geometric features in the bulk perspective, which facilitates their computation -- see~\cite{Almheiri:2019hni,Geng:2020qvw,Chen:2020uac,Chen:2020hmv,Geng:2020fxl,Emparan:2021hyr,Frassino:2022zaz}. 
In this model, the $O(G_N^0)$ quantum corrections to the partition function in the brane picture due to the holographic matter can be computed using semiclassical geometry in the bulk picture. 
The corresponding quantum corrected thermodynamic entropy is equal to the generalised entropy~\cite{Engelhardt:2014gca} between the two asymptotic boundaries. 

We construct semiclassical microstates for this black hole model with two layers of holography, by generalising the microstate construction in~\cite{Balasubramanian:2022gmo}.
We show that in the universality limit where the matter shells are infinitely heavy, the partition function of the microstates can be factorised into a geometric part, corresponding to the quantum corrected partition function of the doubly holographic model~\cite{Grimaldi:2022suv}, and a universal part. 
We compute the dimension of the Hilbert space spanned by such microstates, and find the expected relation to the generalised entropy, including the quantum corrections from the matter degrees of freedom.
This quantum corrected statistical entropy also coincides with the thermodynamic entropy and the generalised entropy between the two asymptotic boundaries of the black holes
\bea
S_{\text{micro}}=S_{\text{thermo}}=S_{\text{gen}}~.
\eea

Interesting future works in this direction include extending the construction to higher-dimensional doubly holographic black holes (for example, the quantum BTZ black hole~\cite{Emparan:2020znc}), and to black holes coupled to more generic matter, which have been discussed in \cite{Climent:2024trz}.

Two sections follow. In \autoref{sec: easy island}, we review the doubly holographic black hole of \cite{Grimaldi:2022suv} and compute the generalised entropy between the left and right asymptotic boundaries.
Then in \autoref{sec: microstates} we construct semiclassical microstates for this black hole, compute their overlap statistics and identify the quantum corrected entropy upon state counting. 
\autoref{app: corner} provides supplementary details on corner terms in the gravitational action for manifolds with piecewise smooth boundaries, which play a role in the construction of black hole microstates.
For convention: We ignore volume elements in integrals, as should be clear from the integration manifold. We use $(g_{\mn},h_{ij},\gamma_{ab})$ to denote the metrics and indices for the bulk, codim$-1$ (branes and matter shells) and codim$-2$ (corner) manifolds.

\section{Black holes with holographic matter}
\label{sec: easy island}

In this section, we summarise the model of~\cite{Grimaldi:2022suv} and discuss some of its properties. This doubly holographic model of a BTZ black hole with a brane can be viewed from three perspectives: the bulk, the brane, and the boundary perspective. 

\begin{figure}[t]
    \centering
    \includegraphics[scale=0.45]{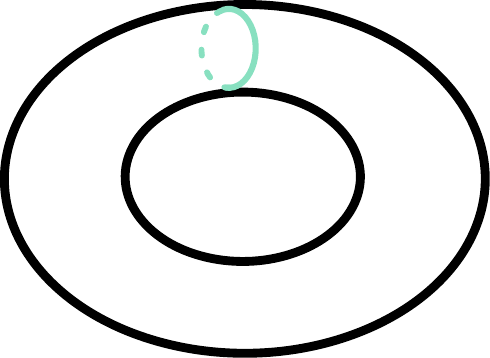} \quad 
    \includegraphics[scale=0.45]{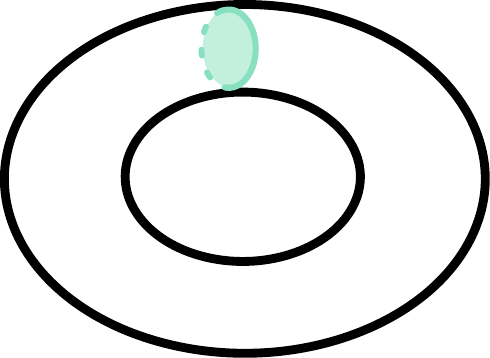} \quad
    \includegraphics[scale=0.45]{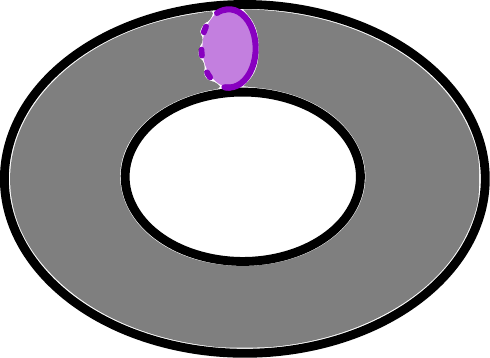} \\ 
    (a) \hspace{3.6cm} (b)\hspace{3.6cm} (c)
    \caption{Illustration of the three perspectives of the model in Euclidean signature. a) In the boundary perspective, there are CFT degrees of freedom on the torus coupled to a conformal defect, depicted in teal. b) In the brane perspective, we dualise the defect degrees of freedom but not those of the ambient CFT. Specifically, the holographic dual to the defect degrees of freedom consists of a gravitating brane with holographic CFT degrees of freedom, also in teal. The brane is coupled to the CFT on the torus. c) In the bulk perspective, the holographic dual to the CFT degrees of freedom on the brane and on the torus correspond to bulk Einstein gravity on the solid torus. There is an intrinsic JT gravity interface brane depicted in purple.} 
    \label{fig:persp}
\end{figure}

\paragraph{Bulk perspective.} We begin from the bulk perspective, in which we have a bulk gravitational theory of Einstein gravity in three dimensions coupled to a two-dimensional brane -- see \autoref{fig:persp}c. We will be interested in the setting in which the brane has an intrinsic gravity, specifically a Jackiw–Teitelboim gravity action. The action is given by\footnote{To illustrate the construction of the model, we omit the relevant Gibbons-Hawking-York terms to have a well-defined variational principle. These terms can be found in eq.~\eqref{eq: I total}. Note that in JT gravity, the dynamics are governed by the Gibbons-Hawking-York term at the asymptotic boundary of the brane~\cite{Almheiri:2014cka,Maldacena:2019cbz}.}
\begin{equation} \label{eq: dyn action}
    I = I_{\rm bulk} + I_{\rm JT} + I_{\rm ct}\,,
\end{equation}
where\footnote{In this work, we focus on the Euclidean continuation of the construction of~\cite{Grimaldi:2022suv}. For this reason, we work in Euclidean signature throughout.}
\begin{equation}\label{eq: action terms}
\begin{aligned}
     I_{\rm bulk}&= -\frac{1}{16 \pi G_N} \int d^3x \sqrt{g}\left[ R+\frac{2}{L^2}\right]\,, \\
     I_{\rm JT} & = - \frac{1}{16\pi G_{\rm brane}} \int d^2x \sqrt{h}\left[\varphi_0 \tilde{R} + \varphi \left( \tilde{R}+ \frac{2}{\ell_{\rm JT}^2}\right)\right]\,,\\
     I_{\rm ct} & = \frac{1}{4\pi G_N L}\int d^2 x \sqrt{h}\,.
\end{aligned}
\end{equation}

The parameters of this theory are the bulk Newton's constant $G_N$ and AdS scale $L$, and the brane Newton's constant $G_{\rm brane}$  and AdS scale $\ell_{\rm JT}$. 
The Ricci scalar associated with the bulk metric $g_{\mu\nu}$ is denoted by $R$, while the induced metric on the brane $h_{ij}$ leads to the induced Ricci scalar on the brane denoted by $\tilde{R}$. 
The dilaton on the brane has a constant part $\varphi_0$ and a varying part $\varphi$. We have included a counterterm $I_{\rm ct}$ in the brane action to ensure that the Karch-Randall induced gravity action corresponds to a gravity theory in AdS$_2$ with length scale $\ell_{\rm JT}$, as will become clear around eq.~\eqref{eq: I induced}.

The bulk metric equations of motion set the bulk geometry to be locally AdS$_3$ with length scale $L$, and the dilaton equation of motion sets the brane geometry to be locally AdS$_2$ with length scale $\ell_{\rm JT}$. The brane metric equation is trivial because the Einstein-Hilbert action is topological in two dimensions. The location of the brane is determined by the Israel junction conditions~\cite{Israel:1966rt}, which read~\cite{Grimaldi:2022suv}
\begin{equation}\label{eq: junction condition}
\Delta K_{ij} - h_{ij} \Delta K = - 2 \sqrt{1-\frac{L^2}{\ell_{\rm JT}^2}} h_{ij} \,,
\end{equation}
where $\Delta K_{ij}$ is the discontinuity of the extrinsic curvature across the brane and $\Delta K = h^{ij} \Delta K_{ij} $ is its trace. Taking the trace of this equation leads to
\begin{equation}
    \Delta K = 4 \sqrt{1-\frac{L^2}{\ell_{\rm JT}^2}}\,.
\end{equation}
Hence, the junction conditions can be equivalently written as
\begin{equation}
    \Delta K_{ij} = \frac{\Delta K}{2} h_{ij}\,, \quad {\rm where} \quad \Delta K = 4 \sqrt{1-\frac{L^2}{\ell_{\rm JT}^2}}\,.
\end{equation}

For the bulk geometry to be a black hole, the asymptotic boundary conditions are those of a BTZ black hole: a torus with radii $R$ and $\frac{\beta}{2\pi}$, respectively.\footnote{To ensure the dominant bulk geometry is a BTZ black hole and not thermal AdS, we restrict the relation between $\beta$ and $R$ so that the temperature $1/\beta$ is above the Hawking-Page temperature~\cite{Hawking:1982dh}. Note that this temperature asymptotes to zero in the limit in which the brane geometry becomes flat $L/\ell_{\rm JT}\ll 1$ -- see Appendix C of~\cite{Grimaldi:2022suv}.} The BTZ geometry that fills these boundary conditions is

\begin{equation}\label{eq: ds BTZ}
    ds^2_{\rm BTZ} = \left(\frac{r^2}{L^2} - \mu^2 \right)\frac{L^2}{R^2} d\tau^2 + \frac{dr^2}{\frac{r^2}{L^2} - \mu^2 } + r^2 d\phi^2\,,
\end{equation}
where $\mu = \frac{2\pi R}{\beta}$. The coordinates $\phi$ and $\tau$ have periodicity $2\pi$ and $\beta$ respectively to avoid any conical singularities. After an appropriate rescaling, the induced metric on the asymptotic boundary $r\to \infty$ is
\begin{equation}
    ds^2_{\rm bdy} = d\tau^2 +R^2 d\phi^2\,,
\end{equation}
which corresponds to the torus where the circumferences of the $\tau-$ and $\phi-$ cycles are $\beta$ and $2\pi R$, respectively.

\begin{figure}
    \centering
    \includegraphics[width=0.8\linewidth]{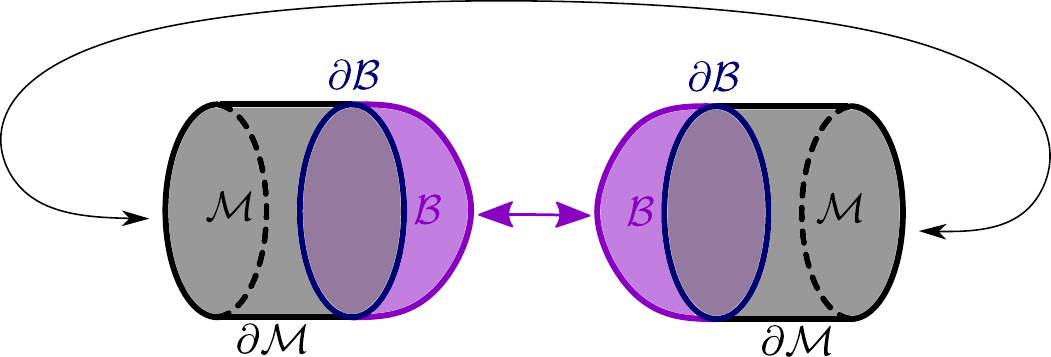}
    \caption{The gluing of the BTZ solution along the brane (purple) in a $Z_2$ symmetric way. Note that the bases of the two grey cylinders are identified. The resulting topology is a solid torus.}
    \label{fig:geombulk}
\end{figure}

The brane is anchored at the asymptotic boundary along the cycle given by $\phi=0$. An important feature of gravity in three dimensions is that the Einstein equations are restrictive enough to impose the bulk geometry to be locally AdS$_3$, so the presence of the brane does not affect the geometry away from its location. The resulting geometry can therefore be constructed by gluing the BTZ solution~\eqref{eq: ds BTZ} in a $Z_2$ symmetric way along the location of the brane, see \autoref{fig:geombulk}, which will be parametrised by a function $f_{\rm brane}$ in the bulk as follows
\begin{equation} \label{eq: brane-equation}
    \phi = f_{\rm brane}(r)\,,
\end{equation}
with the brane being extended in the $\tau$ direction. The function $f_{\rm brane}(r)$ can be found by solving the Israel junction conditions~\eqref{eq: junction condition} and is given by~\cite{Grimaldi:2022suv}
\begin{equation}\label{eq: brane-profile}
    f_{\rm brane}(r) = \frac{1}{\mu} {\rm arcsinh}\left( \frac{k\mu L}{r}\right)\,, \quad {\rm where} \quad k^2 +1=\frac{\ell^2_{\rm JT}}{L^2} \,.
\end{equation}

The induced metric on the brane is a locally AdS$_2$ black hole with the following metric
\begin{equation}\label{eq: BTZ metric}
\begin{aligned}
    ds^2_{\rm brane} &= \left(\frac{r^2}{L^2} -\mu^2\right) \frac{L^2}{R^2}d\tau^2  + \left(\frac{1}{\frac{r^2}{L^2}-\mu^2} + k^2\right) dr^2\\
    & = \left(\frac{\rho^2}{\ell_{\rm JT}^2} -\mu^2\right) \frac{\ell_{\rm JT}^2}{R^2}d\tau^2 + \frac{d\rho^2}{\frac{\rho^2}{\ell_{\rm JT}^2} -\mu^2}\,,
\end{aligned}
\end{equation}
where we have introduced the new radial coordinate $\rho^2 = r^2 + k^2 \mu^2 L^2$ on the brane. The temperature of the two dimensional black hole on the brane is the same as the bulk BTZ solution.

The brane metric equations determine the dilaton profile up to a boundary value of the dilaton $\bar{\varphi}_r$~\cite{Maldacena:2016upp,Grimaldi:2022suv}, and give
\begin{equation}\label{eq: dilaton profile}
    \varphi(\rho) = \frac{G_{\rm brane}}{G_{\rm eff}} + \frac{\bar{\varphi}_r}{\ell_{\rm JT} \mu } \rho\,,
\end{equation}
where 
\begin{equation}
G_{\rm eff} = \frac{G_N}{L} 
\end{equation}
is the effective Newton's constant for the induced gravity theory on the brane, as we will now explain.

\paragraph{Brane perspective.} The backreaction of the brane enlarges the geometry and causes new graviton modes to localise near the brane~\cite{Karch:2000ct,Chen:2020uac,Chen:2020hmv}. Furthermore, the holographic duality allows for an effective description of the brane as a two dimensional gravity theory coupled to two copies of a holographic CFT with finite cutoff -- see~\cite{Chen:2020uac,Chen:2020hmv,Grimaldi:2022suv} for more details. The induced action can be found by expanding the bulk on-shell action (with the addition of an appropriate Gibbons-Hawking-York boundary term) in a Fefferman-Graham expansion around the asymptotic boundary and integrating radially up to the location of the brane~\cite{Randall:1999ee,Randall:1999vf,Karch:2000ct,Chen:2020uac,Chen:2020hmv,Grimaldi:2022suv}. The resulting terms are combined with the intrinsic brane action to yield an induced brane action
\begin{equation}
    I_{\rm induced} = 2I_{\rm diver} + I_{\rm brane}\,,
\end{equation}
which reads
\begin{equation}\label{eq: I induced}
\begin{aligned}
    I_{\rm induced} &= \frac{1}{16\pi G_{\rm eff}} \int \left(\frac{2}{\ell_{\rm JT}^2} + \tilde{R} \log \left(- \frac{L^2}{8}\tilde{R}\right) +\frac{L^2}{8} \tilde{R}^2 + \cdots \right)\\
    & \quad + \frac{1}{16 \pi G_{\rm brane}} \int \left[ \bar{\varphi}_0 \tilde{R} + \varphi \left( \tilde{R}+ \frac{2}{\ell_{\rm JT}^2}\right)\right]\,,
\end{aligned}
\end{equation}
where the topological Einstein-Hilbert part of the induced action is combined with the JT action, shifting the topological term
\begin{equation}
    \bar{\varphi}_0 = \varphi_0 + \frac{G_{\rm brane}}{G_{\rm eff}}\,.
\end{equation}
The brane perspective of this model consists of a holographic CFT on an Euclidean torus coupled to an AdS$_2$ brane anchored at the $\phi=0$ cycle -- \autoref{fig:persp}b. The brane has a gravity theory with action given by~\eqref{eq: I induced} coupled to the same holographic CFT matter as on the torus. 
Importantly, the AdS$_2$ brane geometry consists of a black hole with holographic matter. 
In~\autoref{sec: microstates}, we will build the semiclassical microstates of this two-dimensional black hole, carefully accounting for the contribution from the holographic matter, and determine the correct Hilbert space dimension including the quantum corrections to the Bekenstein-Hawking entropy.

\paragraph{Boundary perspective.} The boundary perspective is found by invoking the holographic dictionary to interchange the AdS$_2$ gravity + CFT degrees of freedom on the brane with its holographic dual, which corresponds to a conformal defect on the $\phi=0$ cycle of the torus in which the holographic CFT is located -- see \autoref{fig:persp}a. The conformal defect has additional quantum mechanical degrees of freedom dual to the JT gravity + CFT theory on the brane. This perspective is useful for defining well-understood quantities in quantum theory because it does not involve any gravity degrees of freedom. This is the way in which entanglement entropy was carefully computed in~\cite{Almheiri:2019hni,Chen:2020uac,Chen:2020hmv,Grimaldi:2022suv}, and provides a natural way of defining a thermofield double state whose overlap is computed by the geometry which we have described in the section above. We will also start from this perspective to define a family of quantum corrected semiclassical black hole microstates in~\autoref{sec: microstates}, following the formalism of~\cite{Balasubramanian:2022gmo,Balasubramanian:2022lnw,Balasubramanian:2024rek}.
Before doing so, we review the thermodynamic and entanglement properties of this model in~\autoref{sec:thermo}.

\subsection{On-shell action and thermodynamics} 
\label{sec:thermo}

The on-shell action and thermodynamic properties of this model have been computed in~\cite{Grimaldi:2022suv}, and we now summarize them here since they will be useful for the microstate counting in~\autoref{sec: microstates}. The total Euclidean action of the geometry in the bulk perspective is
\begin{equation} \label{eq: I total}
    I_{\rm tot} = I_{\rm EH}^{\cal M} + I_{\rm GH}^{\partial {\cal M}} + I_{\rm ct}^{\partial {\cal M}} + I_{\rm JT}^{\cal B} + I_{\rm ct}^{\cal B} + 2 I_{\rm GH}^{\cal B} + I_{\rm GH}^{\partial {\cal B}} + I_{\rm ct}^{\partial {\cal B}}\,.
\end{equation}
The bulk Einstein-Hilbert action $I_{\rm EH}^{\cal M}$, the JT brane action $I_{\rm JT}^{\cal B}$ and the brane counterterm $I_{\rm ct}^{\cal B}$ were already present in eq.~\eqref{eq: dyn action}. Because the bulk and brane have dynamical gravity actions, we also have to include the appropriate boundary Gibbons-Hawking-York terms
\begin{equation}
    I_{\rm GH}^{\partial \cal M} = -\frac{1}{8\pi G_N}\int_{\partial {\cal M}} K\,, \quad I_{\rm GH}^{\cal B} = -\frac{1}{8\pi G_N}\int_{{\cal B}} K\,, \quad I_{\rm GH}^{ \partial \cal B} = -\frac{1}{8\pi G_{\rm brane}}\int_{\partial {\cal B}} \tilde{K}\,,
\end{equation}
where $K$ is the trace of the extrinsic curvature of the bulk codimension one surfaces $\partial \cal M$ and $\cal B$, and $\tilde{K}$ is the trace of the extrinsic curvature of the boundary of the brane $\partial {\cal B} = {\cal B} \cap \partial \cal M$. Additionally, we add counterterms $I_{\rm ct}^{\partial {\cal M}}$ and $I_{\rm ct}^{\partial {\cal B}}$ at the asymptotic boundary to ensure the total on-shell action is finite
\begin{equation}
    I_{\rm ct}^{\partial {\cal M}} = \frac{1}{8\pi G_N L} \int_{\partial {\cal M}}1\,, \quad I_{\rm ct}^{\partial {\cal B}}= \frac{1}{8\pi G_{\rm brane} \ell_{\rm JT}} \int_{\partial {\cal B}} \varphi_b\,.
\end{equation}
For details of the computation, see appendix C of~\cite{Grimaldi:2022suv}. Adding all the contributions together, the total Euclidean on-shell action is
\begin{equation}\label{eq: action-island-total}
    I_{E} = -\frac{\varphi_0+\bar{\varphi}_r}{4G_{\rm brane}} - \frac{\pi^2 R}{2G_N \beta} - \frac{1}{2 G_N}{\rm arcsinh}{(k)}\,.
\end{equation}

From the brane perspective, the first term in~\eqref{eq: action-island-total} is the leading semiclassical on-shell action of the JT gravity theory living on the brane. 
The second term is due to the holographic CFT degrees of freedom coupled to the JT gravity on the brane, and can be considered a quantum correction to the semiclassical result. 
The last term is due to the conformal defect which sources the brane at the asymptotic boundary, and is associated with the boundary degrees of freedom. We will call the totality of~\eqref{eq: action-island-total} the quantum corrected Euclidean action, and the corresponding partition function
\begin{equation}\label{eq: partition function}
Z(\beta) = e^{-I_E}\,,
\end{equation}
will be referred to as the quantum corrected partition function. The energy is given by
\begin{equation}\label{eq: E beta}
    E = -\partial_\beta \log Z(\beta) = \partial_\beta I_E = \frac{\pi^2 R}{2 G_N \beta^2} \,,
\end{equation}
which is entirely due to the quantum corrections from the CFT degrees of freedom on the brane. 

Using that the free energy is given by $F=I_E/\beta$, the thermodynamic entropy of this system is
\begin{equation}\label{eq: entropy-tot}
    S = -\frac{d F}{dT} = -\frac{d\beta}{dT} \frac{dF}{d\beta} = \frac{\varphi_0+\bar{\varphi}_r}{4G_{\rm brane}} +  \frac{\pi^2 R}{G_N\beta} + \frac{1}{2G_N} {\rm arcsinh}(k)\,.
\end{equation}
Once again, from the brane perspective, the first term corresponds to the thermodynamic entropy of the JT gravity degrees of freedom, while the second and third term correspond to the entropy of the holographic CFT degrees of freedom and the conformal defect, respectively. From the brane perspective, the latter two terms can be interpreted as quantum corrections to the leading semiclassical result given by the value of the dilaton at the horizon. This interpretation will be further supported by computing the entanglement entropy between the two asymptotic boundaries, in which the generalised entropy includes the quantum corrections to the leading Bekenstein-Hawking entropy in the brane perspective. 

Lastly, we can also work in the microcanonical ensemble, where we fix the energy $E$ instead of the inverse temperature $\beta$. In this case, the microcanonical entropy is found by inverting the relation between energy and temperature~\eqref{eq: E beta} and plugging it into~\eqref{eq: entropy-tot},
\begin{equation}\label{eq: mircro-entropy-tot}
    \mathbf{S} = -\frac{d F}{dT} = -\frac{d\beta}{dT} \frac{dF}{d\beta} = \frac{\varphi_0+\bar{\varphi}_r}{4G_{\rm brane}} +  \sqrt{\frac{2G_N E}{\pi^2 R}} + \frac{1}{2G_N} {\rm arcsinh}(k)\,.
\end{equation}
As will be shown in~\autoref{sec: microstates}, the dimension of the Hilbert space spanned by the semiclassical black hole microstates is related to the quantum corrected microcanonical entropy~\eqref{eq: mircro-entropy-tot}.

\subsection{Entanglement entropy}

Upon analytic continuation to Lorentzian time $t=i \tau$, the resulting geometry corresponds to a double-sided BTZ black hole with a JT brane connecting the two asymptotic boundaries. 
We take the $t=0$ slice and compute the entanglement entropy between the two asymptotic boundaries. We would like to compare the thermodynamic entropy~\eqref{eq: entropy-tot} with the entanglement entropy between the two asymptotic boundaries. 
To do this, we use the prescription for holographic entanglement entropy including quantum corrections~\cite {Engelhardt:2014gca}. 
From the brane perspective, it is given by the generalised entropy associated with the quantum extremal surface $\sigma_{\mathbf{R}}$ on the brane, which is a codimension two surface (a point), homologous to one of the asymptotic boundaries $\mathbf{R}$ and which minimizes the generalised entropy
\begin{equation}\label{eq: Sgen}
    S_{\rm gen} = S_{\rm BH} + S_{\rm CFT}\,.
\end{equation}
The generalised entropy contains two terms, the first one is the semiclassical Bekenstein-Hawking entropy for the JT gravity theory on the brane
\begin{equation}\label{eq: SBH}
    S_{\rm BH} = \frac{\varphi_0+\varphi (\sigma_{\mathbf{R}})}{4 G_{\rm brane}}\,,
\end{equation}
and the second one consists of quantum corrections, given by the entanglement entropy of the CFT degrees of freedom on the brane between $\sigma_\mathbf{R}$ and the asymptotic boundary $\mathbf{R}$. These can be computed for non holographic CFTs as in~\cite{Almheiri:2019psf}. When the CFT matter on the brane is holographic, its entanglement entropy is given by the RT prescription
\begin{equation}\label{eq: SCFT}
    S_{\rm CFT} = \frac{{\rm Length}(\Sigma_{\mathbf{R}})}{4G_N}\,.
\end{equation}
The RT surface $\Sigma_\mathbf{R}$ in eq.~\eqref{eq: SCFT} is the extremal surface homologous to $\mathbf{R}$ with minimum length.

As has been emphasised in~\cite{Chen:2020uac,Chen:2020hmv,Grimaldi:2022suv}, this rule follows naturally from the bulk perspective, in which it is given by the usual holographic entanglement entropy prescription with the addition of a contact term due to the presence of gravitating degrees of freedom on the brane. Concretely, the entanglement entropy of any boundary subregion $\mathbf{R}$ is given by
\begin{equation} \label{eq: SEE}
    S_{\rm EE}(\mathbf{R}) = \min_{\partial\Sigma_{\mathbf{R}}'=\partial\mathbf{R} \cup \sigma_{\mathbf{R}}'}\left[ \frac{{\rm Length}(\Sigma_{\mathbf{R}}')}{4G_N} + \frac{\varphi_0 + \varphi(\sigma_{\mathbf{R}}')}{4G_{\rm brane}} \right]\,,
\end{equation}
where the minimization is taken over candidate QES $\sigma_\mathbf{R}'$ and candidate RT surfaces $\Sigma_\mathbf{R}'$. The remarkable feature in doubly holographic models is that the quantum corrections~\eqref{eq: SCFT} in the brane perspective are given by a purely geometric quantity in the bulk perspective, and correspond to the semiclassical Bekenstein-Hawking entropy of the BTZ black hole, which is dual to the holographic matter on the brane.

\begin{figure}
    \centering
    \includegraphics[scale=0.8]{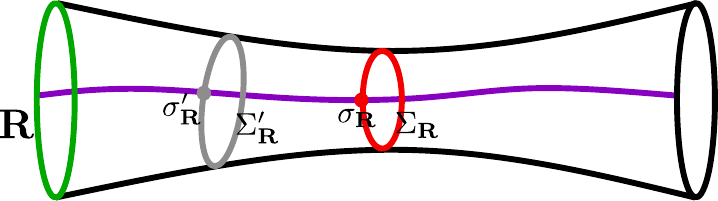} 
    \caption{A constant time slice of the Euclidean semiclassical geometry in \autoref{fig:geombulk} with the brane in purple. The subregion $\mathbf{R}$ is taken to be one of the two boundaries visualised in green. The candidate extremal surfaces $\Sigma'_\mathbf{R}$ are indicated in grey and intersect the brane at $\sigma'_\mathbf{R}$. The minimal surface $\Sigma_\mathbf{R}$ given in red coincides with the horizon and intersects the brane at $\sigma_\mathbf{R}$.}
    \label{fig:EER}
\end{figure}

In this case, we consider the boundary region $\mathbf{R}$ to be one of the two asymptotic boundaries, and in particular, it has no boundary $\partial \mathbf{R} = \varnothing$. Therefore, the candidate RT surfaces $\Sigma_{\mathbf{R}}'$ correspond to closed one-dimensional surfaces that wind once around the $\phi-$cycle -- see~\autoref{fig:EER}. These candidate RT surfaces intersect the brane at a specific location $\sigma_\mathbf{R}'$, and because of time translation symmetry, the entirety of the extremizing RT surface $\Sigma_{\mathbf{R}}$ will be at the $t=0$ slice, and so we can restrict our minimization to surfaces that are time independent. For every intersection $\sigma_\mathbf{R}'$, there is a single geodesic winding around the $\phi-$cycle with minimal length corresponding to the distance between the point and its image in a covering space of the BTZ geometry. The distance between two points with coordinates $(t_i,r_i,\phi_i)$, $i=1,2$ in the BTZ geometry~\eqref{eq: ds BTZ} is given by~\cite{Grimaldi:2022suv}
\begin{equation}
    \cosh{\frac{d}{L}} = \frac{r_1 r_2}{\mu^2 L^2 } \cosh \mu(\phi_1-\phi_2) - \frac{\sqrt{(r_1^2-\mu^2L^2) (r_2^2-\mu^2L^2)}}{\mu^2 L^2} \cosh \frac{\mu (t_1-t_2)}{R}\,.
\end{equation}
To apply this formula for a point $\sigma_\mathbf{R}'$ on the brane and its image point, we use
\begin{equation}
    t_1=t_2=0\,, \quad r_1=r_2=r\,, \quad \phi_1= f_{\rm brane}(r)\,, \quad \phi_2 = -2\pi - f_{\rm brane}(r)\,,
\end{equation}
so that the length of the minimal candidate RT surface $\Sigma_\mathbf{R}'$ intersecting the brane at some radius $r$ is
\begin{equation}\label{eq: length Sigma}
    \cosh \frac{{\rm Length}(\Sigma_{\mathbf{R}}')}{L} =  1+ \frac{r^2}{\mu^2 L^2} \left(\cosh\left( 2 \pi \mu + 2\, {\rm arcsinh} \left(\frac{k \mu L}{r}\right)\right) -1\right)\,.
\end{equation}

Combining~\eqref{eq: length Sigma} with the profile of the dilaton~\eqref{eq: dilaton profile} and recalling the relation between the radii $\rho^2 = r^2 + k^2 \mu ^2 L^2$, we can show that the generalised entropy~\eqref{eq: SEE} is monotonic in $r$, or equivalently in $\rho$, since
\begin{equation}
\begin{aligned}
    \partial_\rho  \frac{{\rm Length}(\Sigma_{\mathbf{R}}')}{4G_N}   & = \frac{ \sqrt{2} L \sinh\left(\pi \mu\right)}{2G_N \sqrt{2 \mu^2 L^2 + r^2 \cosh\left( 2 \mu \left(f_{\rm brane}(r)+\pi\right)\right)-r^2 }} \,, \\
   \partial_\rho \frac{\varphi_0 + \varphi(\sigma_{\mathbf{R}}')}{4G_{\rm brane}} & = \frac{\bar{\varphi}_r}{4G_{\rm brane} \ell_{\rm JT} \mu} \,,
\end{aligned}
\end{equation}
are both manifestly positive. The RT surface minimizing~\eqref{eq: SEE} is therefore that for which the intersection with the brane occurs at the smallest possible radius, namely the horizon $r_h=\mu L$. The length of the RT surface is
\begin{equation}
    {\rm Length} \left( \Sigma_\mathbf{R} \right) = 2 \pi \mu L + 2L \, {\rm arcsinh} \left( k\right)\,.
\end{equation}
Note that this length is exactly equal to the area of the event horizon $A_{\rm horizon} = 2 r_h\int_{-\pi}^{f(r_h)} d\phi$. The value of the dilaton at the QES can be found using eq.~\eqref{eq: dilaton profile} evaluated at $\rho=\mu \ell_{\rm JT}$
\begin{equation}
    \varphi\left(\sigma_\mathbf{R}\right) = \frac{G_{\rm brane}}{G_{\rm eff}} + \bar{\varphi}_r\,.
\end{equation}

The quantum extremal surface $\sigma_\mathbf{R}$ minimizing the generalised entropy~\eqref{eq: Sgen} is the event horizon on the brane, and the bulk RT surface $\Sigma_\mathbf{R}$ extremizing the entanglement entropy of the CFT degrees of freedom outside of the black hole is the bulk event horizon. The leading semiclassical contribution to the generalised entropy is therefore
\begin{equation}
    S_{\rm BH} = \frac{\bar{\varphi}_0 + \bar{\varphi}_r}{4G_{\rm brane}}\,,
\end{equation}
and the quantum correction from the holographic CFT degrees of freedom is
\begin{equation}
    S_{\rm CFT}  =  \frac{L}{G_N}\frac{\pi^2 R}{\beta} + \frac{L}{2G_N} {\rm arcsinh}(k)\,.
\end{equation}
Together, we find that the entanglement entropy between the two asymptotic boundaries~\eqref{eq: Sgen} exactly matches the thermodynamic entropy of the system~\eqref{eq: entropy-tot}.

\section{Microstates for black holes with holographic matter}
\label{sec: microstates}
In this section, we construct the microstates for the black holes on the brane with holographic matter described in \autoref{sec: easy island} using the formalism developed in \cite{Balasubramanian:2022gmo,Balasubramanian:2022lnw}.
Then we perform the state counting and find that the dimension of the Hilbert space spanned by these states is the exponential of the quantum corrected entropy.

\subsection{Microstates for black holes with holographic matter}

\begin{figure}
    \centering
    \includegraphics[width=0.5\linewidth]{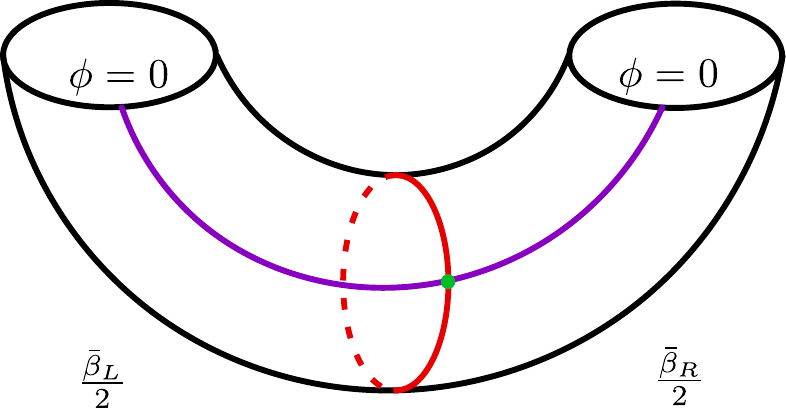}
    \caption{The Euclidean path integral that prepares the state given in \eqref{eq: state-CFT-gen}. The spherically symmetric shell is denoted by the red circle, which intersects with the conformal defect at $\phi=0$ denoted by the purple line.}
    \label{fig:state}
\end{figure}

\begin{figure}
    \centering
    \includegraphics[scale=0.38]{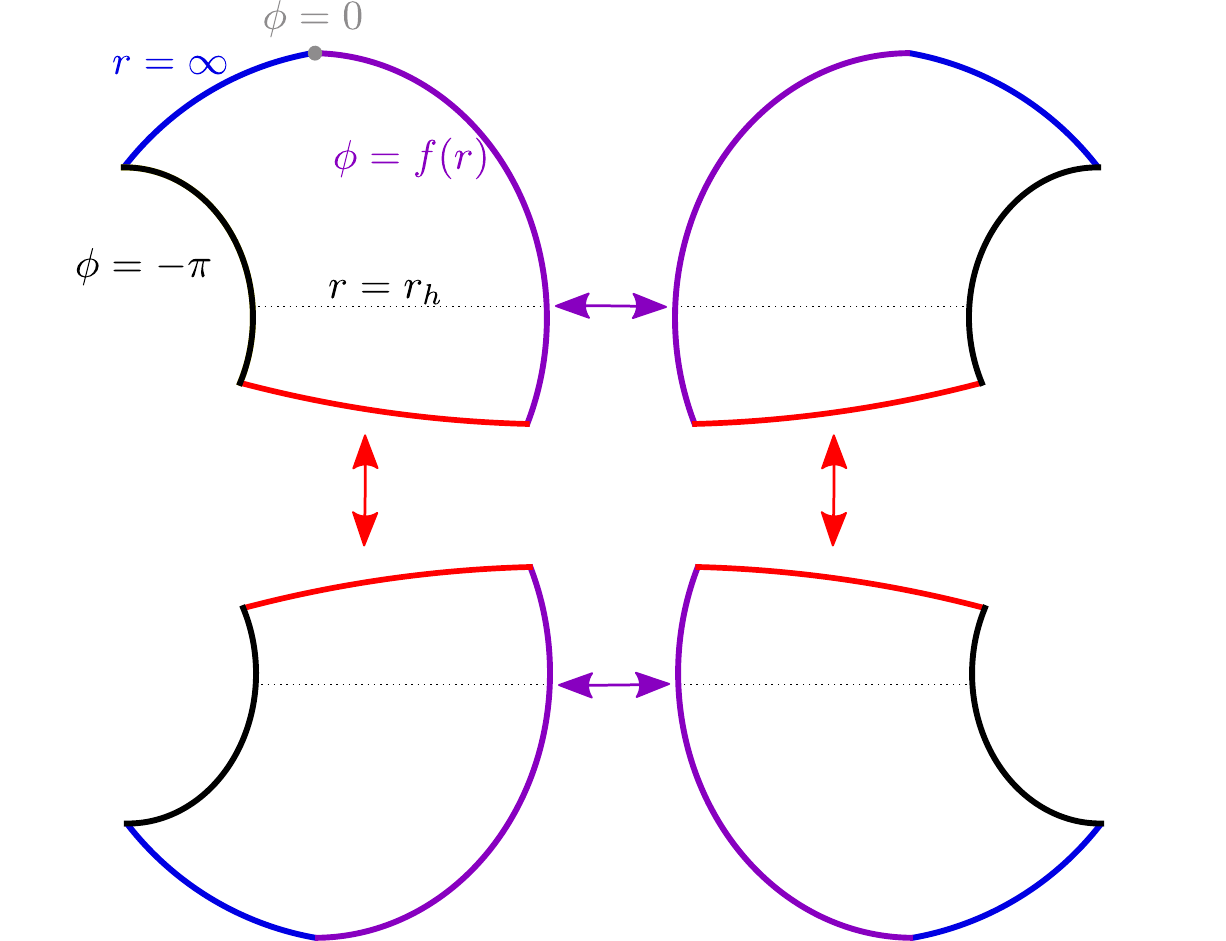} \quad 
    \includegraphics[scale=0.5]{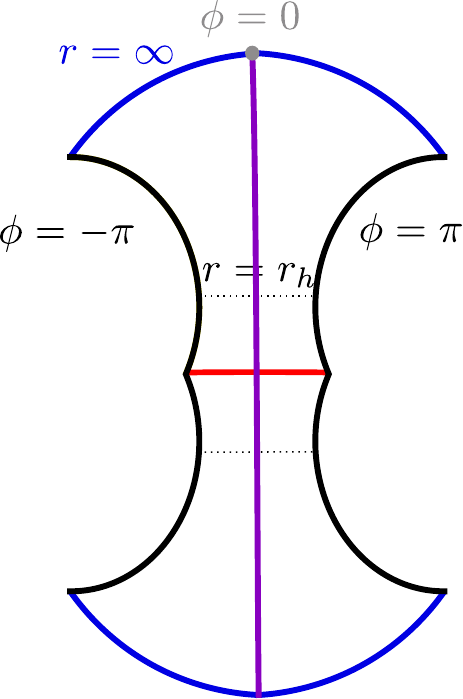} \\ 
    \hspace{1.3cm} (a) \hspace{6.7cm} (b)
    \caption{A constant time slice of the Euclidean semiclassical geometry dual to the state given by \eqref{eq: state-CFT-gen}. The thin shell of matter dual to the operator $\cO^{(k)}$ is denoted by red, and the brane connecting the two asymptotic boundaries is denoted by purple. The time slice is identified at the $\phi=-\pi$ and $\phi=\pi$ curves.  Left: the bulk geometry is cut by the shells and branes, and the identification for the gluing. Right: the geometry after the gluing. }
    \label{fig:geom}
\end{figure}

\begin{figure}
    \centering
    \includegraphics[width=0.6\linewidth]{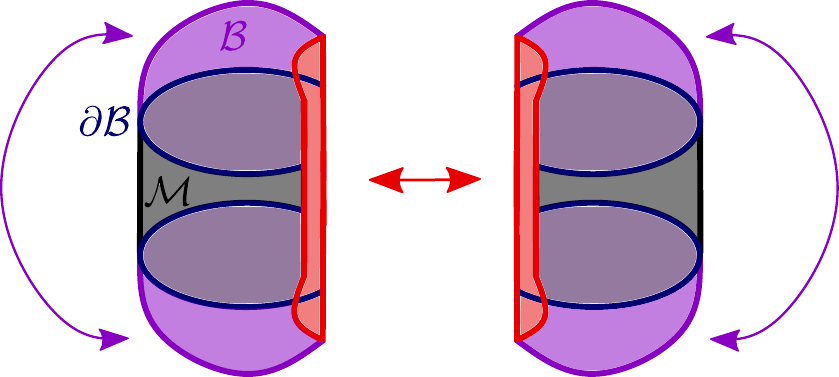}
    \caption{The 3d Euclidean semiclassical geometry dual to the state given by \eqref{eq: state-CFT-gen}. The brane $\cB$ is denoted by the purple region, and the thin shell is denoted by the red region. The brane trajectory is given by \eqref{eq: brane-equation} and the shell trajectory is given by \eqref{eq: trajectory-shell}, along which the geometry is identified as indicated by the arrows. The resulting topology is a solid torus.}
    \label{fig:geomshell}
\end{figure}

We consider states defined on a tensor product of two copies of a $2d$ defect CFT (dCFT) with the topology of a cylinder, which is the boundary picture of the doubly holographic model introduced in \autoref{sec: easy island}.
Each dCFT has a Hamiltonian and energy basis $H\ket{m}=E_m\ket{m}$.
We construct a family of states (\autoref{fig:state}) by inserting operators $\cO^{(k)}$ that are dual to spherically symmetric thin shells of matter with some corresponding mass $m_k$, and then evolving by the Hamiltonian to the right and left over Euclidean times $\bar{\beta}_R/2$ and $\bar{\beta}_L/2$.
This yields normalised states of the form
\bea\label{eq: state-CFT-gen}
|\Psi_k\rangle =\frac{1}{\sqrt{Z_{\Psi}}} \sum_{n, m} e^{-\frac{1}{2} \bar{\beta}_L E_m-\frac{1}{2} \bar{\beta}_R E_n} \mathcal{O}^{(k)}_{m n}|m\rangle_L \otimes|n\rangle_R~,
\eea
where $\mathcal{O}^{(k)}_{m n}=\langle m| \mathcal{O}^{(k)}|n\rangle$ and $Z_{\Psi}=\operatorname{Tr}\left[\mathcal{O}^{(k)\dagger} e^{-\bar{\beta}_L H} \mathcal{O}^{(k)} e^{-\bar{\beta}_R H}\right]$, and the trace is taken over a single copy of the CFT. The states constructed in this way are dual to semiclassical spatial wormholes connecting two asymptotically $\ads_3$ regions with a brane stretched between two boundaries, similar to the easy island model \cite{Grimaldi:2022suv} reviewed in \autoref{sec: easy island}, see \autoref{fig:geom} and \autoref{fig:geomshell}, but with the addition of the spherically symmetric thin shell which extends the wormhole and intersects the brane.

\begin{figure}
    \centering
    \includegraphics[scale=0.78]{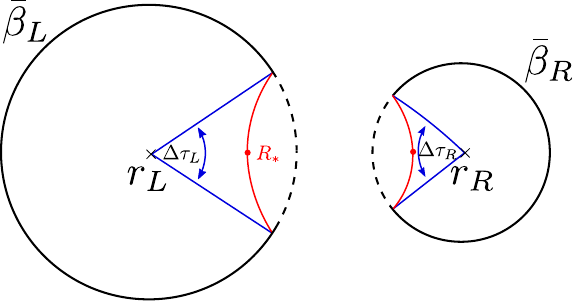} \quad 
    \includegraphics[scale=0.7]{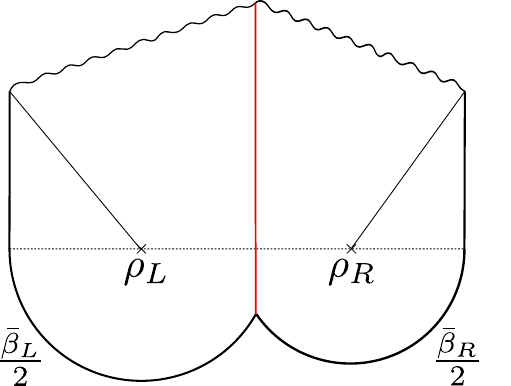} \\ 
    \hspace{1.3cm} (a) \hspace{6.7cm} (b)
    \caption{Left: The trajectory (red line) of the 
    thin shell in a constant $\phi$ slice of the Euclidean $\ads_3$, given
by \eqref{eq: trajectory-shell}.
The horizons $r_L$ and $r_R$ are denoted by the crosses.
The ``$R$'' and ``$L$'' regions are glued along the shell trajectory. The dashed curves and the shell trajectory enclose the excised part of the spacetime. Right: the Lorentzian continuation of the Euclidean geometry. The shell is behind the horizon.
}
    \label{fig:def}
\end{figure}

The entire bulk spacetime computing the norm $Z_{\Psi}$ can be thought of as composed of four pieces resulting from the intersection of the time-translation invariant brane and the spherically symmetric shell.
To know the geometry and compute the on-shell action, we need to know the trajectory of the brane (eq.~\eqref{eq: brane-equation} and eq.~\eqref{eq: brane-profile}) and the shell, respectively, and how they intersect.
The trajectory of the shell of matter with mass $m_k$ can be parameterized by $(\tau(T), r(T))$ where $T$ is the proper time on the shell, which
is also determined by the Israel junction condition \cite{Israel:1966rt,Balasubramanian:2022gmo}, see \autoref{fig:def}
\bea\label{eq: trajectory-shell}
r(T)&=R_*\cosh T~,\\
\tau_{L,R}(r)&=\frac{1}{r_{L,R}}\tan^{-1}\left(\frac{r_{L,R}\sqrt{r^2-R_*^2}}{r\sqrt{R_*^2-r_{L,R}^2}}\right)~,
\eea
where $R_*=\sqrt{r_R^2+\left(\frac{r^2_{R}-r^2_{L}}{8G_N m_k}-2 G_N m_k\right)^2}$ is the turning point of the trajectory. The horizons of the left and right BTZ black holes are located at $r_{L,R}=\mu_{L,R}L$. We refer to the intersection of the brane and the shell as the corner $\cal C$, a one dimensional line, whose trajectory is given by eq.~\eqref{eq: brane-equation} and eq.~\eqref{eq: trajectory-shell}.
There is an angular deficit at the corner, so the manifold is not smooth.
This can be seen by computing the angle between the unit normals to the brane and the shell worldvolume.
The unit normals can be determined from the trajectories \eqref{eq: trajectory-shell} and \eqref{eq: brane-equation}
\bea
&\text{brane:}~& n_{\mu}= \left(0, \frac{k}{\sqrt{1 + k^2} \, r}, \sqrt{\frac{r^2 + k^2 r_{L,R}^2}{1 + k^2}}\right)~,\\
&\text{shell:}~&n'^{\mu}=\left({-\frac{\sqrt{r^2 - R_*^2}}{r^2 - r_{L,R}^2},\sqrt{R_*^2 - r_{L,R}^2},0}\right)~.
\eea
The angle $\theta_{L,R}$ between the two trajectories is given by
\bea
\cos \theta_{L,R}=n\cdot n'=n_r n'^r=\frac{k\sqrt{R_*^2 - r_{L,R}^2} }{\sqrt{1 + k^2} \, r}=\frac{k \, \sqrt{R_*^2 - r_{L,R}^2} \, \text{sech}(T)}{\sqrt{1 + k^2} \, R_*}~,
\eea
and the total defect angle $\Delta\Theta$ is
\bea\label{eq: defect-angle}
\Delta\Theta=2\pi-2(\theta_R +\theta_L)~,
\eea
which enters into the corner term \eqref{eq: corner-action} in the on shell action~\cite{Hayward:1993my,Lehner:2016vdi}, when the total geometry has a conical defect, as is explained in~\autoref{app: corner}.
We will focus on the universality limit $m_k\to\infty$, in which the auxiliary quantities due to the operator insertion in \eqref{eq: state-CFT-gen} become universal (independent from the geometric quantities of the original black holes)
\bea
R_* &\to 2 G_N m_k~,\\
\theta_{L,R}&\to \cos^{-1}\frac{k \,\sech(T)}{\sqrt{k^2+1}}~,\\
\Delta\Theta&\to 4\sin^{-1}\frac{k \,\sech(T)}{\sqrt{k^2+1}}~.
\eea

Now we are ready to construct the on-shell actions $I_{\text{tot}}$ for the geometry which computes the norm of the state \eqref{eq: state-CFT-gen}  ($Z_{\Psi}=e^{-I_{\text{tot}}}$), in the universality limit
\bea
I_{\text{tot}}=I_{L}^{\text{ren}}+I_R^{\text{ren}} +I_{\text{uni.}}^{\text{ren}}~,
\eea
where we directly write down renormalized actions with possible GHY terms and counter terms included, denoted by "ren".
The renormalized on-shell actions $I_{L,R}^{\text{ren}}$ are those of the BTZ black hole with a brane, given by \eqref{eq: action-island-total} with inverse temperatures $\beta_{L,R}$ for the left and right black holes.
The universal temperature independent term is given by
\bea
I_{\text{uni.}}=I_{\text{shell}}+I_{\cC}+I_{\text{shell}/\cC}~,
\eea
which appears due to the presence of the shell of matter.
The action of the shell is \cite{Balasubramanian:2022gmo}
$I_{\text{shell}}= \int_{\cS}\sigma$, and the corner term $I_{\cC}$ at the intersection of the brane and the shell is given by \eqref{eq: corner-action} with the defect angle \eqref{eq: defect-angle}.
This intersection also sources a shell of matter on the brane $I_{\text{shell}/\cC}$, whose stress tensor satisfies $T_{ab}\propto\Delta \Theta \gamma_{ab}$, as explained below \eqref{eq: corner-action}.
We do not give the explicit expressions in this work, as they drop out in the state counting procedure due to the normalization of the states.

\subsection{State counting and microscopic entropy}\label{sec: counting}
Having obtained the geometry and on-shell actions dual to the microstates \eqref{eq: state-CFT-gen}, we are ready to perform the state counting and give the quantum-corrected microscopic entropy for the black hole on the brane.
We will follow the state counting procedure developed in~\cite{Balasubramanian:2022gmo,Balasubramanian:2022lnw}, which we briefly summarise and highlight the new ingredients due to the use of quantum-corrected microstates. 
Firstly, we project the states in \eqref{eq: state-CFT-gen} onto the microcanonical band $\left[E_{L, R}, E_{L, R}+\delta E\right)$ with a projector $\Pi_E=\Pi_E^L \otimes \Pi_E^R$
\begin{equation}
\left|\psi_k^E\right\rangle=\frac{1}{\sqrt{\left\langle\Psi_k\right| \Pi_E\left|\Psi_k\right\rangle}} \Pi_E\left|\Psi_k\right\rangle~,
\end{equation} 
Next, we consider the space spanned by $\Omega$ of these microstates in the limit where each state has a shell with infinite mass $m_k\to\infty$
\begin{equation}\label{eq:Kspanmicro2}
    {\cal H}^E_{\rm bulk}(\Omega) \equiv {\rm Span}\{|\psi^E_k\rangle, \quad k=1,\ldots,\Omega\}~.
\end{equation}
When $\Omega$ is large enough, the states in the span become linearly dependent, and the dimension saturates to the dimension of the black hole Hilbert space. This can be diagnosed through the kernel of the Gram matrix $G$ whose entries are given by
\begin{equation}
G_{i j}=\left\langle\Psi_i^E \mid \Psi_j^E\right\rangle~.
\end{equation}
The aim is to determine the value of $\Omega$ at which $G$ first develops a zero eigenvalue and the kernel is no longer empty. To this end, we introduce the resolvent of this Gram matrix
\begin{equation}\label{eq: resolvent-expansion}
R_{i j}(\lambda):=\left(\frac{1}{\lambda \mathbb{I}-G}\right)_{i j}=\frac{1}{\lambda} \delta_{i j}+\sum_{n=1}^{\infty} \frac{1}{\lambda^{n+1}}\left(G^n\right)_{i j}~. 
\end{equation}
The trace of this matrix, $R(\lambda)$ has poles at each eigenvalue of $G$, and the residue of each pole counts the degeneracy of the corresponding eigenvalue. Of particular interest is the value of $\Omega$ at which $R(\lambda)$ develops a pole for $\lambda=0$. Further increasing $\Omega$ increases the degeneracy of the zero eigenvalue in such a way that the number of linearly independent states, and therefore the rank of the Gram matrix remains unchanged.
The trace of the resolvent $R(\lambda)$ can be computed using the gravitational path integral as follows:
\begin{itemize}
    \item The partition function of n-boundary wormholes is given by the semiclassical approximation to the gravitational path integral 
    \bea
    \overline{Z^n}=Z\left(n \beta_L\right) Z\left(n \beta_R\right) e^{-n I_{\text {univ }}}~,
    \eea
    where $Z\left( \beta\right)=e^{-I(\beta)}$ is the quantum corrected partition function of the $\ads_2$ black hole on the brane, given by \eqref{eq: action-island-total}.
    \item After projecting to the microcanonical window with an inverse Laplace transform, the microcanonical partition function is
    \bea
\overline{\mathbf{Z}^n}&=\frac{\mathbf{h}_n}{2 \pi} \int d q_L d q_R e^{n q_L E_L+n q_R E_R} Z\left(n q_L\right) Z\left(n q_R\right) e^{-n I_{\text {univ }}}\\
&=e^{\mathbf{S}_L+\mathbf{S}_R-n I_{\text {univ }}}~,
    \eea
where $\mathbf{h}_n$ is the Hessian determinant of $-\log Z(nq_L) Z(nq_R)$ with respect to $q_{L,R}$ evaluated at the saddle point. The quantum corrected microcanonical entropies $\mathbf{S}_{L,R}=\mathbf{S}(E_{L,R})$ are given by~\eqref{eq: mircro-entropy-tot}.
\item Focusing on the planar limit where both $\Omega$ and $e^{1/G_N}$ are large, the series expansion of the resolvent \eqref{eq: resolvent-expansion} simplifies \cite{Penington:2019kki} 
\begin{equation}\label{eq: resolvent-planar}
\overline{R_{i j}(\lambda)}=\frac{1}{\lambda} \delta_{i j}+\frac{1}{\lambda} \sum_{n=1}^{\infty} \frac{\overline{\mathbf{Z}^n}}{\overline{\mathbf{Z}}^n} \overline{R(\lambda)}^{n-1} \overline{R_{i j}(\lambda)}~,
\end{equation}
where the universal temperature independent contribution $I_{\text{univ}}$ drops out of the equation.
Taking the trace leads to a quadratic equation that is solved by\footnote{There are two solutions to this quadratic equation, and the correct solution can be identified from the fact that the gram matrix has $\Omega$ non-negative real eigenvalues and its trace equals $\Omega$.}
\begin{equation}
\overline{R(\lambda)}=\frac{e^{\mathbf{S}_L+\mathbf{S}_R}(\lambda-1)+\Omega + \sqrt{\left(e^{\mathbf{S}_L+\mathbf{S}_R}(\lambda-1)+\Omega\right)^2-4 e^{\mathbf{S}_L+\mathbf{S}_R} \lambda \Omega}}{2 \lambda}~.
\end{equation}
\item As $\lambda\rightarrow 0$, the solution becomes
\begin{equation}
\overline{R}=\frac{\Omega-e^{\mathbf{S}_L+\mathbf{S}_R}}{\lambda} \Theta\left(\Omega-e^{\mathbf{S}_L+\mathbf{S}_R}\right)+\dots~,
\end{equation}
where $\dots$ denotes terms regular in $\lambda$.
As $\Omega<e^{\mathbf{S}_L+\mathbf{S}_R}$, $\overline{R}$ is regular at $\lambda=0$ and there are no zero-eigenvalues.
When $\Omega>e^{\mathbf{S}_L+\mathbf{S}_R}$, the trace of the resolvent develops a residue $\operatorname{Res}_{\lambda=0} \bar{R}=\Omega-e^{\mathbf{S}_L+\mathbf{S}_R}$, which corresponds to the number of zero-eigenvalues.
Therefore, $\overline{\operatorname{Rank}G}=\operatorname{min}\{\Omega, e^{\mathbf{S}_L+\mathbf{S}_R}\}$, and the dimension of the black hole Hilbert space is $e^{\mathbf{S}_L+\mathbf{S}_R}$.
\end{itemize}

To conclude, using the state counting procedure developed in~\cite{Balasubramanian:2022gmo,Balasubramanian:2022lnw}, we find that the dimension of the black hole Hilbert space is given by the exponential of the quantum corrected microcanonical entropy~\eqref{eq: mircro-entropy-tot}. The new ingredients here are the quantum corrected partition function, which we obtained from classical geometry using double holography.
As a result, the entropy obtained by state counting is the same as the thermodynamical entropy and the generalised entropy between the two asymptotic boundaries.

\section*{Acknowledgments}
We thank Vijay Balasubramanian, Ben Craps and Tom Yildirim for helpful discussions.
We would like to especially thank Andrew Svesko for the helpful discussion that motivated this study.
Work at VUB was supported by FWO-Vlaanderen project G012222N and by the VUB Research Council through the Strategic Research Program High-Energy Physics. JH is supported by FWO-Vlaanderen through a Junior Postdoctoral Fellowship.


\appendix
\section{Corner term in Euclidean signature}\label{app: corner}

In this work, we are concerned with spacetimes that have thin shells of matter intersecting branes, which results in conical defects along a codimension-two surface which consists of the intersection of the two. These geometries have been explicitly built by gluing patches of black hole geometries with various boundaries, connected along joints that we refer to as corners. When the boundary of a geometry has corners, the gravitational action requires the addition of corner terms to have a well-defined Dirichlet variational problem~\cite{Hayward:1993my,Lehner:2016vdi}. This implies that the total action for geometries with conical defects constructed by gluing various such corner manifolds together also requires a similar term along the defect surface. In this appendix, we adapt the analysis of~\cite{Hayward:1993my,Lehner:2016vdi} to the Euclidean manifolds and comment on the junction conditions on the shells and corners in the microstate geometries of the main text.

Consider a spacetime $\cM$ with piecewise-smooth boundaries $\cB_I$ ($I=1,2,\dots$).
The gravitational action with the GHY term on the boundary is
\bea
I&=I_{EH}+I_{GHY}\\
&=-\frac{1}{2\kappa_N}\int_{\cM} (R-2\Lambda)-\frac{1}{\kappa_N}\int_{\cB_I}K~.
\eea
The variation with the Dirichlet boundary condition on $\cB_I$ is
\bea
\delta I=-\frac{1}{2\kappa_N}\int_{\cM} (G_{\mu\nu}+\Lambda g_{\mu\nu})\delta g^{{\mu\nu}}-\frac{1}{2\kappa_N}\int_{\cB_I}\dslash v^{\mu} n_{\mu}~,
\eea
where $\dslash$ denotes a non-exact variation.
The second term on the r.h.s. is only present when the boundaries $\cB_I$ have boundaries themselves, 
\bea 
\dslash v^{\mu} n_{\mu}&=-D_i \dslash A^i, & \dslash A^i&=-e^i_{\mu}n_{\nu}\delta g^{\mu\nu}~,
\eea
where $D$ is the covariant derivative on $\cB_I$.
The vielbein between the bulk and the brane is denoted by $e^{\mu}_i$, and its inverse is given by 
$e^i_{\mu}=g_{\mn}h^{ij}e^{\nu}_j$ where $g_{\mn}$ and $h^{ij}$ are the metrics of the bulk and the brane.
Using Gauss's law, the integral on $\cB_I$ can be further written as integrals on the boundaries of the boundaries (corners),
\bea
\int_{\cB}\dslash v^{\mu} n_{\mu}=-\int_{\cB_{I 2}}\dslash A^i r_{i2} +\int_{\cB_{I 1}}\dslash A^i r_{i1}~,
\eea
where $\cB_{I2}$ is the outer corner, $\cB_{I 1}$ is the inner corner and $r_{1,2}$ is the outward pointing unit normal vector on the corners.

\begin{figure}
    \centering
    \includegraphics[width=0.5\linewidth]{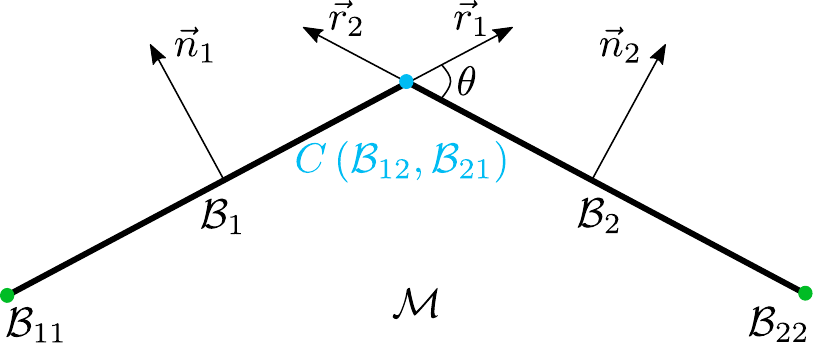}
    \caption{The corner $\cC$ formed by the non-smooth intersection of piecewise-smooth boundaries $\cB_1$ and $\cB_2$ which results in a defect angle $\theta$. The normal vectors  of the boundaries $\cB_{1,2}$ are $\vec{n}_{1,2}$ and $\vec{r}_{1,2}$ are the normal vectors of the corners $\cB_{12}$ and $\cB_{21}$.}
    \label{fig:corner}
\end{figure}

When the corners are joint by two boundaries, the two corner terms combine into a closed variation.
Consider a corner $\cC$ joint by $\cB_1$ to the left and $\cB_2$ to the right so that $\cC=\cB_{12}=\cB_{21}$ (\autoref{fig:corner}). Ignoring other corners of $\cB_I$, the boundary term becomes
\bea\label{eq: corner from boundary}
\int_{\cB_1+\cB_2}\dslash v^{\mu} n_{\mu}&=\int_{\cC} \dslash C~, 
\eea
where
\bea
\dslash C&=\dslash A_2^i r_{i2}-\dslash A_1^i r_{i1}\\
&=(r_2^{\mu} n_2^{\nu}-r_1^{\mu} n_1^{\nu})\delta g_{\mu\nu}~.
\eea
Using the geometric relation
\begin{align}
  n_2&=\cos\theta\, n_1+\sin \theta\, r_1~, & r_2&=-\sin\theta\, n_1+\cos \theta\, r_1~,  
\end{align}
one can express the normal vectors $r_{1,2}$ on the corner by the normal vectors $n_{1,2}$ on $\cB_{1,2}$ and the angle $\theta$ between them. From the definition of the angle $\cos \theta=g^{\mu\nu}n_{1\mu}n_{2\nu}$, one can also relate the $\delta \theta$ with $\delta g_{\mu\nu}$. In the end, we obtain a simple relation between the corner term $\dslash C$ and the angle $\theta$ \cite{Hayward:1993my,Lehner:2016vdi}
\bea
\dslash C=-2\delta \theta
\eea
Therefore, in order to have a well-defined variation problem with Dirichlet boundary condition, one needs to add another corner term $I_{\cC}$ to the action to cancel the contribution from \eqref{eq: corner from boundary},
\bea\label{eq: corner-action}
I&=I_{EH}+I_{GHY}+I_C~,\\
I_{\cC} &=-\frac{1}{\kappa_N}\int_{\cC} \theta~.
\eea
Finally, let us comment on the junction conditions on the piecewise-smooth thin shells with corners in the interior of the spacetime, where no Dirichlet boundary condition is imposed.
The Israel junction condition can be derived from the variational problem on the shells.
On the corner, the variation $\delta I_{\cC}$ produces another term proportional to $\int_{\cC}\theta\gamma^{ab}\, \delta \gamma_{ab}$ where $\gamma_{ab}$ is the induced metric on the corner, which has to be balanced by some matter on the corner \cite{Hayward:1993my}.

\bibliographystyle{JHEP}
\bibliography{microstates.bib}

\end{document}